\documentclass[aps,prl,twocolumn,preprintnumbers,showpacs,showkeys,nofootinbib]{revtex4-1}
\usepackage{latexsym, amsfonts, amssymb, amsmath, mathrsfs, mathtools}

\newcommand{\C}{\mathbb{C}}

\newcommand{\ii}{\mathrm{i}}

\begin{document}

\preprint{ZMP-HH/16-7}

\title{Elliptic solid-on-solid model's partition function as a single determinant}

\author{W. Galleas}
\email[E-mail address: ]{wellington.galleas@desy.de}
\thanks{The work of W.G. is supported by the German Science Foundation (DFG) under the Collaborative Research Center (SFB) 676: Particles, Strings and the Early Universe.}

\affiliation{II. Institut f\"ur Theoretische Physik, Universit\"at Hamburg, Luruper Chaussee 149, 22761 Hamburg, Germany.}

\date{\today}

\begin{abstract}
In this work we express the partition function of the integrable elliptic solid-on-solid model with domain-wall boundary conditions as a single determinant. This representation
appears naturally as the solution of a system of functional equations governing the model's partition function.
\end{abstract}

\pacs{05.50+q, 02.30.IK}
\keywords{Elliptic integrable systems, domain-wall boundaries, functional equations}

\maketitle


\section{Introduction} \label{intro}

Several types of lattice models are \emph{exactly solvable} in the sense that the summation defining their partition 
function can be expressed as a closed formula without any approximation. This is usually a highly non-trivial task but it has been achieved
for certain models enjoying the \emph{gift} of integrability \cite{Baxter_book}. Two-dimensional lattice models are rather special and among
them we find the most notorious exactly solved models of Statistical Mechanics. For instance, the Ising model and the $8$-vertex model. These models 
are corner stones of the modern theory of integrable systems and, in particular, a series of developments were due to Baxter's ingenious
works on the $8$-vertex model \cite{Baxter_1973a, Baxter_1973b, Baxter_1973c}. In the course of studying eigenvectors of the symmetric
$8$-vertex model Baxter has introduced the so called \emph{solid-on-solid} models or \textsc{sos} models for short. They are also refereed to
in the literature as \emph{interaction-round-a-face} (\textsc{irf}) models and they differ from vertex models in the way lattice interactions
are characterized. While vertex models assign configuration variables to the edges of a rectangular lattice, \textsc{sos} models associate
configuration variables with lattice sites. In this way, the \textsc{sos} model dual to Baxter's $8$-vertex model consists of an Ising type
model with four-spin interaction as discussed in \cite{Baxter_book}.

Boundary conditions are among the main ingredients when defining a lattice statistical system and the elliptic \textsc{sos} model with 
domain-wall boundaries has received special attention recently. This special type of boundary conditions were firstly introduced by Korepin for the
$6$-vertex model \cite{Korepin_1982} and subsequently translated to \textsc{sos} models in \cite{Korepin_Justin_2000}. Interestingly, for this
particular type of boundary conditions the models' partition functions can be written down explicitly as a closed formula \cite{Izergin_1987, Rosengren_2009, Galleas_2013},
in contrast to the case with periodic boundary conditions. For the latter the \emph{solution} still relies on the resolution of Bethe ansatz equations \cite{Lieb_1967}.

The $6$-vertex model with domain-wall boundaries has found several applications, ranging from enumerative combinatorics 
\cite{Kuperberg_1996} to the study of gauge theories \cite{Szabo_2012}, and the elliptic \textsc{sos} model is not far behind. 
A series of works have been devoted to the study of its combinatorial properties \cite{Rosengren_2011} and relation to special polynomials 
\cite{Rosengren_2015}. These results are mainly due to Rosengren's representation \cite{Rosengren_2009} for the model's partition function as a sum
of Frobenius type determinants which seem to generalize Izergin's single determinant representation for the $6$-vertex model \cite{Izergin_1987}.

Although a compact expression for the elliptic \textsc{sos} model's partition function has been found in \cite{Galleas_2013}, the
possibility of expressing such partition function as a single determinant has eluded the researchers of the field so far. This is not
only of interest for the computation of physical quantities but providing a definitive answer to this puzzle also shed new light onto
the mathematical structure underlying elliptic integrable systems. This is precisely the purpose of this letter and in what 
follows we show how a single determinant representation can be derived from the analysis of special functional relations
originated from the \emph{dynamical Yang-Baxter algebra}.

\section{The model} \label{MODEL}

Write $\mathscr{L}_n \coloneqq \{ 1, 2, \dots , n \}$ and let $(i,j) \in \mathscr{L}_{L+1} \times \mathscr{L}_{L+1}$ be $2$-tuples describing a two-dimensional
square lattice. Hence our lattice is formed by the juxtaposition of $L \times L$ square cells which we shall simply refer to as \emph{faces}. We assign a statistical
weight $w_{ij}$ to the face enclosed by the Cartesian coordinates $(i,j)$, $(i,j+1)$, $(i+1,j)$ and $(i+1,j+1)$. The configuration of a given face $w_{ij}$
is characterized by variables $\{ h_{i,j}, h_{i,j+1}, h_{i+1,j}, h_{i+1,j+1}  \}$ and the system's partition function is defined as
\begin{equation} \label{PF}
Z \coloneqq \sum_{\{ h_{i,j} \}} \prod_{i,j=1}^{L+1} w_{ij} \begin{pmatrix} h_{i+1,j} & h_{i+1,j+1} \\ h_{i,j} & h_{i,j+1} \end{pmatrix} \; .
\end{equation}
The variable $h_{i,j}$ is also referred to as \emph{height function} and here $h_{i,j} \coloneqq \tau + n_{i,j} \gamma$ with
$\tau, \gamma \in \mathbb{C}$ and $n_{i,j} \in \mathbb{Z}$. Also, here we consider that $h_{i,j}$ and $h_{i',j'}$ at neighboring sites
can only differ by $\pm \gamma$. The set $\{ h_{i,j} \}$ then contains height functions of allowed face configurations. Baxter's
elliptic \textsc{sos} model has six allowed face configurations and the respective statistical weights are given by,
\begin{eqnarray} \label{BW}
w_{ij} \begin{pmatrix} \tau \pm \gamma & \tau \\ \tau & \tau \mp \gamma \end{pmatrix} &=& [x + \gamma]  \nonumber \\
w_{ij} \begin{pmatrix} \tau \pm \gamma & \tau \pm 2\gamma \\ \tau & \tau \pm \gamma \end{pmatrix} &=& [\tau \pm \gamma] \frac{[x]}{[\tau]} \nonumber \\
w_{ij} \begin{pmatrix} \tau \pm \gamma & \tau \\ \tau & \tau \pm \gamma \end{pmatrix} &=&  [\tau \pm x] \frac{[\gamma]}{[\tau]} \; ,
\end{eqnarray}
where $[x] \coloneqq \frac{1}{2} \sum_{n = -\infty}^{+ \infty} (-1)^{n-\frac{1}{2}} p^{(n+\frac{1}{2})^2} e^{-(2n+1) x}$
for $x \in \mathbb{C}$ and fixed elliptic nome  $0 < p < 1$. The function $[x]$ corresponds to the Jacobi theta-function $\Theta_1 (\ii x, \nu)$
with $p = e^{\ii \pi \nu}$ according to the conventions of \cite{Whittaker_Watson_book}. In order to completely define the partition function $Z$
we also need to declare the boundary conditions being used. Here we shall consider boundary conditions of domain-wall type which corresponds to
the assumptions $h_{1,j} = h_{j,1} = \tau + (L+1-j)\gamma$ and $h_{L+1,j} = h_{j,L+1} = \tau + (j-1)\gamma$.

\section{Algebraic-functional framework} \label{AF}

The algebraic structure underlying the statistical weights \eqref{BW} are nowadays well known. It consists of the elliptic quantum group
$\mathcal{E}_{p, \gamma} [ \widehat{\mathfrak{gl}_2} ]$, as described in \cite{Felder_1994, Felder_1995}, and this enables the so called \emph{dynamical
Yang-Baxter algebra} to be used in the study of the partition function $Z$. Here we shall adopt the procedure developed in \cite{Galleas_2013}
which exploits the dynamical Yang-Baxter algebra as a source of functional equations characterizing quantities of physical interest. 
We now write $Z = Z_{\tau} (x_1, x_2 , \dots , x_L)$ in order to capture the dependence of our partition function with the relevant variables. 
The variables $x_i \in \mathbb{C}$ will be referred to as \emph{spectral parameters} while $\tau$ will be called \emph{dynamical parameter}. In addition to that
$Z$ also depends on \emph{inhomogeneity parameters} $\mu_i \in \mathbb{C}$ ($ 1 \leq i \leq L$) and an \emph{anisotropy parameter} $\gamma \in \mathbb{C}$.
The latter are fixed from now on. Using the algebraic-functional framework we have shown in \cite{Galleas_2013} that the partition function
\eqref{PF} satisfies the following functional equation,
\begin{equation} \label{eqA}
M_0 \; Z_{\tau} (X) + \sum_{i \in \{0,1, \dots, L \}} N_i \; Z_{\tau + \gamma} (X_i^0) = 0 \; ,
\end{equation}
where $X \coloneqq \{ x_i \in \C \mid 1 \leq i \leq L \}$ and $X_i^{\alpha} \coloneqq X \cup \{ x_{\alpha} \} \backslash \{ x_i \}$.
The coefficients in \eqref{eqA} explicitly read
\begin{eqnarray} \label{coeffA}
M_0 &\coloneqq& \frac{[\tau + \gamma]}{[\tau + (L+1)\gamma]} \prod_{j=1}^{L} [x_0 - \mu_j]  \\
N_0 &\coloneqq& -\frac{[\tau + 2\gamma]}{[\tau + (L+2)\gamma]} \prod_{j=1}^{L} [x_0 - \mu_j + \gamma] \prod_{j=1}^{L} \frac{[x_j - x_0 + \gamma]}{[x_j - x_0]}  \nonumber \\
N_i &\coloneqq& \frac{[\tau] [\tau + 2\gamma + x_0 - x_i]}{[\tau + (L+2)\gamma][x_i - x_0]} \prod_{j=1}^{L} [x_i - \mu_j + \gamma] \nonumber \\
&& \qquad \qquad \quad \times \prod_{\substack{j=1 \\ j \neq i}}^L \frac{[x_j - x_i + \gamma]}{[x_j - x_i]} \qquad i = 1,2, \dots , L \; . \nonumber 
\end{eqnarray}
We refer to \eqref{eqA} as equation \emph{type A} due to its roots within the algebraic-functional method  \cite{Galleas_2013}.
Using the same method we can also derive an equation of \emph{type D} reading
\begin{eqnarray} \label{eqD}
\bar{M}_0 \; Z_{\tau + \gamma} (X) + \sum_{i \in \{ \bar{0}, 1, \dots, L \}} \bar{N}_i \; Z_{\tau} (X_i^{\bar{0}}) = 0 \; ,
\end{eqnarray}
with coefficients defined as
\begin{eqnarray} \label{coeffD}
\bar{M}_0 &\coloneqq& \prod_{j=1}^{L} [x_{\bar{0}} - \mu_j + \gamma] \nonumber \\
\bar{N}_0 &\coloneqq& - \prod_{j=1}^{L} [x_{\bar{0}} - \mu_j] \prod_{j=1}^{L} \frac{[x_{\bar{0}} - x_j + \gamma]}{[x_{\bar{0}} - x_j]}  \\
\bar{N}_i &\coloneqq&  \frac{[\gamma] [\tau + (L+1)\gamma + x_{\bar{0}} - x_i]}{[x_{\bar{0}} - x_i] [\tau + (L+1)\gamma]} \prod_{j=1}^{L} [x_i - \mu_j] \nonumber \\
&& \qquad \qquad \quad \times \prod_{\substack{j=1 \\ j \neq i}}^L \frac{[x_i - x_j + \gamma]}{[x_i - x_j]} \qquad i = 1,2, \dots , L \; . \nonumber 
\end{eqnarray}
Although each equation \eqref{eqA} and \eqref{eqD} can individually determine the partition function \eqref{PF}, as shown in \cite{Galleas_2013},
here we shall demonstrate how a determinant representation for $Z_{\tau}$ follows from a particular combination of equations type A and D.

\section{Determinant representation} \label{DET}

In the recent paper \cite{Galleas_2016} we have shown how functional equations with structure similar to \eqref{eqA} and \eqref{eqD}
can be solved in terms of determinants. However, in order to tackle our present equations, namely \eqref{eqA} and \eqref{eqD}, we need to generalize the
mechanism of \cite{Galleas_2016}.
The reason for that is the dependence of \eqref{eqA} and \eqref{eqD} with the dynamical parameter $\tau$. Hence we shall look for a suitable combination
of our equations such that $\tau$ no longer plays the role of variable. Now consider \eqref{eqA} under permutations $x_0 \leftrightarrow x_l$ for $0 \leq l \leq L$. This operation leaves us with a set of $L+1$ equations involving
$Z_{\tau} (X^0_l)$ and $Z_{\tau+\gamma} (X^0_i)$ for $0 \leq i \leq L$. Thus we can solve the resulting system of equations and, in particular, express
$Z_{\tau+\gamma} (X)$ as a combination of terms $Z_{\tau} (X^0_l)$. By substituting the result of this procedure in \eqref{eqD} we are then left with the functional relation
\begin{equation} \label{eqAD}
\mathcal{M}_0 \; Z_{\tau} (X) + \sum_{i=1}^L \mathcal{N}_i \; Z_{\tau} (X_i^0) + \sum_{i=1}^L \bar{\mathcal{N}}_i \; Z_{\tau} (X_i^{\bar{0}}) = 0 \; .
\end{equation}
The coefficients of \eqref{eqAD} explicitly read
\begin{eqnarray}
\label{coeffADneu}
\mathcal{M}_0 &\coloneqq& \prod_{j=1}^L \frac{[x_0 - x_j + \gamma] [x_0 - \mu_j] [x_{\bar{0}} - \mu_j + \gamma]}{[x_0 - x_j] [x_0 - \mu_j + \gamma]} \nonumber \\
&& - \prod_{j=1}^L \frac{[x_{\bar{0}} - x_j + \gamma] [x_{\bar{0}} - \mu_j]}{[x_{\bar{0}} - x_j ]} \nonumber \\
\mathcal{N}_i &\coloneqq& - \frac{[\gamma] [x_0 - x_i + \tau + (L+1)\gamma]}{[\tau + (L+1)\gamma] [x_0 - x_i]} \nonumber \\
&& \quad \times \prod_{j=1}^L \frac{[x_i - \mu_j] [x_{\bar{0}} - \mu_j + \gamma]}{[x_0 - \mu_j + \gamma]} \prod_{\substack{j=1 \\ j \neq i}}^L \frac{[x_i - x_j + \gamma]}{[x_i - x_j]} \nonumber \\
\bar{\mathcal{N}}_i &\coloneqq& \frac{[\gamma] [x_{\bar{0}} - x_i + \tau + (L+1)\gamma]}{[\tau + (L+1)\gamma] [x_{\bar{0}} - x_i]} \prod_{j=1}^L [x_i - \mu_j] \nonumber \\
&&  \times \prod_{\substack{j=1 \\ j \neq i}}^L \frac{[x_i - x_j + \gamma]}{[x_i - x_j]} \; .
\end{eqnarray}
Compared to equations \eqref{eqA} and \eqref{eqD}, we can readily see that $\tau$ no longer plays the role of variable and we fix it from this point on.
Next we notice that permutations $x_0 \leftrightarrow x_l$ for $1 \leq l \leq L$, $x_{\bar{0}} \leftrightarrow x_m$ for $1 \leq m \leq L$
and simultaneous permutations $x_0 \leftrightarrow x_l$, $x_{\bar{0}} \leftrightarrow x_m$ for $1 \leq l < m \leq L$ yields new equations with extra terms 
of the form $Z_{\tau} ( X_{i,j}^{0, \bar{0}} )$ where  $X_{i,j}^{\alpha, \beta} \coloneqq X \cup \{x_{\alpha}, x_{\beta} \} \backslash \{x_i, x_j \}$,
in addition to the ones already present in \eqref{eqAD}. The precise form of these new equations are not enlightening at the moment but it is important to remark
that they form a closed system containing one term $Z_{\tau} (X)$, $L$ terms of type $Z_{\tau} (X_i^0)$, $L$ terms of type $Z_{\tau} (X_i^{\bar{0}})$ and $\frac{L(L-1)}{2}$ terms of form
$Z_{\tau} ( X_{i,j}^{0, \bar{0}} )$. On the other hand, we have $L$ equations produced by permutations $x_0 \leftrightarrow x_l$ ($1 \leq l \leq L$),
another $L$ equations obtained from $x_{\bar{0}} \leftrightarrow x_m$ ($1 \leq m \leq L$) and $L(L-1)/2$ originated from the simultaneous
permutations $x_0 \leftrightarrow x_l, x_{\bar{0}} \leftrightarrow x_m$ for $1 \leq l < m \leq L$. Thus we can solve the resulting system of equations
and express each element $Z_{\tau} (X_i^0)$, $Z_{\tau} (X_i^{\bar{0}})$ and $Z_{\tau} ( X_{i,j}^{0, \bar{0}} )$ in terms of $Z_{\tau} (X)$. The proportionality
factor will be a ratio of determinants according to Cramer's rule. For instance, using this approach we can write $Z_{\tau} (X_i^0) = \left( \text{det}(A_i) / \text{det}(B) \right) Z_{\tau} (X)$
with given matrices $A_i$ and $B$ of dimension $d_L \coloneqq L(L+3)/2$. Although the matrices $A_i$ and $B$ exhibit local dependence with the variable
$x_{\bar{0}}$, the ratio $\text{det}(A_i) / \text{det}(B)$ is globally independent of $x_{\bar{0}}$ since it corresponds to $Z_{\tau} (X_i^0) /Z_{\tau} (X)$.
Interestingly, this same feature has already made its appearance in a different context. This is precisely the property allowing for the 
computation of path integrals through the \emph{localization method} \cite{Duistermaat_Heckman_1982} (see also \cite{Szabo_book} for a review).
Therefore, we can choose $x_{\bar{0}}$ at our convenience and the inspection of the entries of $A_i$ and $B$ shows significant simplifications
with the choice $x_{\bar{0}} = \mu_1 - \gamma$. Next we proceed with separation of variables and notice that the ratio $\text{det}(B) / \text{det}( \left. B \right|_{\tau = -\gamma})$ 
is independent of $x_0$. Hence, we can conclude that $Z_{\tau} (X) \sim \text{det}(B) / \text{det}( \left. B \right|_{\tau = -\gamma})$
and again the variable $x_0$ can be chosen at our will \cite{Galleas_tba}. Here we shall fix $x_{0} = \mu_1 - 2\gamma$ and the proportionality factor can be obtained from
the asymptotic behavior presented in \cite{Galleas_2013}.

Following the above described procedure we are left with the solution
\begin{widetext}
\begin{eqnarray} \label{Z}
Z_{\tau} (X) &=& (-1)^L \left(\frac{[(L+1)\gamma]}{[\tau + (L+2)\gamma]}\right)^{d_{L-1}} \left(\frac{[\tau + (L+1)\gamma]}{[L\gamma]}\right)^{d_L} \prod_{i,j = 1}^L [x_i - \mu_j] \prod_{k=1}^L \frac{[k \gamma]}{[\tau + k \gamma]} \nonumber \\
&& \times \frac{[\sum_{l=1}^L (x_l - \mu_l) + (L+1)\gamma ]}{[\sum_{l=1}^L (x_l - \mu_l) + \tau + (L+2)\gamma ]} \text{det} \left( \Omega \; \Omega_{red}^{-1} \right) \; ,
\end{eqnarray}
\end{widetext}
where $\Omega_{red} \coloneqq \left. \Omega \right|_{\tau = - \gamma}$. In its turn the matrix $\Omega$ can be conveniently depicted
as
\begin{equation}
\Omega = \begin{pmatrix} \mathcal{F} & \mathcal{I} & \mathcal{G} \\ \bar{\mathcal{I}} & \mathcal{K} & \mathcal{J} \\ \bar{\mathcal{F}} & \bar{\mathcal{J}} & \bar{\mathcal{G}} \end{pmatrix} \; ,
\end{equation}
where $\mathcal{F}$, $\bar{\mathcal{F}}$, $\mathcal{G}$ and $\bar{\mathcal{G}}$ are sub-matrices of dimension $L\times L$,
$\mathcal{I}$ and $\bar{\mathcal{J}}$ are of dimension $L \times L(L-1)/2$, $\bar{\mathcal{I}}$ and $\mathcal{J}$ are of dimension $L(L-1)/2 \times L$,
while the dimension of $\mathcal{K}$ is $L(L-1)/2 \times L(L-1)/2$. The matrices $\mathcal{F}$, $\bar{\mathcal{F}}$ and $\mathcal{G}$
are diagonal with non-null entries given by
\begin{eqnarray}
\label{FFbG}
\mathcal{F}_{a,a} &\coloneqq&  [2\gamma] \prod_{k=2}^L [\mu_1 - \mu_k - \gamma]  \prod_{\substack{k=1 \\ k \neq a}}^L \frac{[x_k - \mu_1]}{[x_k - \mu_1 + \gamma]}   \nonumber \\
\bar{\mathcal{F}}_{a,a} &\coloneqq& \frac{[\tau + L\gamma]}{[\tau + (L+1)\gamma]} \prod_{k=1}^L [x_a - \mu_k + \gamma] \nonumber \\
&& \qquad\qquad \qquad \qquad  \times  \prod_{\substack{k=1 \\ k \neq a}}^L \frac{[x_k - \mu_1]}{[x_k - \mu_1 + \gamma]} \nonumber \\
\mathcal{G}_{a,a}  &\coloneqq& \frac{[\tau + (L+2) \gamma]}{[\tau + (L+1)\gamma]} \prod_{k=1}^L [\mu_1 - \mu_k - 2\gamma] \nonumber \\
&& \qquad\qquad \qquad \qquad  \times \prod_{\substack{k=1 \\ k \neq a}}^L \frac{[x_k - \mu_1 + \gamma]}{[x_k - \mu_1 + 2\gamma]}  . \nonumber \\
\end{eqnarray}
On the other hand, the matrix $\bar{\mathcal{G}}$ is a full-matrix with entries
\begin{eqnarray}
\label{Gb}
\bar{\mathcal{G}}_{a,b} \coloneqq \begin{cases}
- \frac{[x_a - \mu_1 + 2 \gamma]}{[x_a - \mu_1 + \gamma]} \prod_{k=1}^L [x_a - \mu_k] \prod_{\substack{k=1 \\ k \neq a}}^L \frac{[x_a - x_k + \gamma]}{[x_a - x_k]} \\
\hfill  b = a \nonumber \\
\frac{[\gamma] [x_a - x_b + \tau + (L+1)\gamma]}{[\tau + (L+1)\gamma] [x_a - x_b]} \frac{[x_b - \mu_1 + 2\gamma]}{[x_b - \mu_1 + \gamma]} \prod_{k=1}^L [x_b - \mu_k] \\
\hfill \times \prod_{\substack{k=1 \\ k \neq a,b}}^L \frac{[x_b - x_k + \gamma]}{[x_b - x_k]} \qquad \text{otherwise}
\end{cases} . \\
\end{eqnarray}
As for the remaining matrices it is convenient to introduce an index $n \colon \mathbb{Z}_{> 0} \times \mathbb{Z}_{> 0} \to \mathbb{Z}_{> 0}$.
More precisely, we define $n_{r,s} \coloneqq s + L (r-1) - \frac{r(r+1)}{2}$ for $1 \leq r < s \leq L$, and in this way we have
\begin{eqnarray}
\label{I}
\mathcal{I}_{a, n_{r,s}} \coloneqq \begin{cases}
\frac{[\gamma] [\mu_1 - x_s + \tau + L\gamma] [\mu_1 - x_s - 3\gamma]}{[\tau + (L+1)\gamma] [\mu_1 - x_s - \gamma] [\mu_1 - x_s - 2\gamma]} \\
\hfill \times \prod_{k=1}^L [x_s - \mu_k] \prod_{\substack{k=1 \\ k \neq r,s}}^L \frac{[x_s - x_k + \gamma]}{[x_s - x_k]} \qquad a = r \nonumber \\
\frac{[\gamma] [\mu_1 - x_r + \tau + L\gamma] [\mu_1 - x_r - 3\gamma]}{[\tau + (L+1)\gamma] [\mu_1 - x_r - \gamma] [\mu_1 - x_r - 2\gamma]} \\
\hfill \times \prod_{k=1}^L [x_r - \mu_k] \prod_{\substack{k=1 \\ k \neq r,s}}^L \frac{[x_r - x_k + \gamma]}{[x_r - x_k]} \qquad a = s \nonumber \\
0 \hfill \text{otherwise} 
\end{cases} \\
\end{eqnarray}
and 
\begin{eqnarray}
\label{Jb}
\bar{\mathcal{J}}_{a, n_{r,s}} \coloneqq \begin{cases}
\frac{[\gamma] [\mu_1 - x_s + \tau + (L-1)\gamma]}{[\tau + (L+1)\gamma] [x_s - \mu_1 + \gamma]} \prod_{\substack{k=1 \\ k \neq r,s}}^L \frac{[x_s - x_k + \gamma]}{[x_s - x_k]}  \\
\hfill \times \prod_{k=1}^L \frac{[x_a - \mu_k + \gamma] [x_s - \mu_k]}{[\mu_1 - \mu_k - \gamma]} \qquad a = r \nonumber \\
\frac{[\gamma] [\mu_1 - x_r + \tau + (L-1)\gamma]}{[\tau + (L+1)\gamma] [x_r - \mu_1 + \gamma]} \prod_{\substack{k=1 \\ k \neq r,s}}^L \frac{[x_r - x_k + \gamma]}{[x_r - x_k]}  \\
\hfill \times \prod_{k=1}^L \frac{[x_a - \mu_k + \gamma] [x_r - \mu_k]}{[\mu_1 - \mu_k - \gamma]} \qquad a = s \nonumber \\
0 \hfill \text{otherwise}
\end{cases}. \\
\end{eqnarray}
Next we turn our attention to the matrices $\bar{\mathcal{I}}$ and $\mathcal{J}$. The entries of $\bar{\mathcal{I}}$ are then given by
\begin{eqnarray}
\label{Ib}
\bar{\mathcal{I}}_{n_{l,m} , b} \coloneqq \begin{cases}
\frac{[2 \gamma] [x_m - \mu_1 + \tau + (L+2)\gamma]}{[\tau + (L+1)\gamma] [x_m - \mu_1 + \gamma]}  \prod_{k=1}^L [\mu_1 - \mu_k - \gamma] \\
\hfill \times \prod_{\substack{k=1 \\ k \neq l, m}}^L \frac{[x_k - \mu_1]}{[x_k - \mu_1 + \gamma]} \qquad  b=l \nonumber \\
- \frac{[2 \gamma] [x_l - \mu_1 + \tau + (L+2)\gamma]}{[\tau + (L+1)\gamma] [x_l - \mu_1 + \gamma]} \prod_{\substack{k=1 \\ k \neq l, m}}^L \frac{[x_k - \mu_1]}{[x_k - \mu_1 + \gamma]}  \\
\hfill \times \prod_{k=1}^L \frac{[\mu_1 - \mu_k - \gamma] [x_m - \mu_k + \gamma]}{[x_l - \mu_k + \gamma]} \qquad b=m \\
0 \hfill \text{otherwise} 
\end{cases} , \\
\end{eqnarray}
while $\mathcal{J} \coloneqq \mathbf{0}$ is a null-matrix. Lastly, we have the following expression for the entries of $\mathcal{K}$,
\begin{widetext}
\begin{eqnarray}
\label{K}
\mathcal{K}_{n_{l,m} , n_{r,s}} \coloneqq \begin{cases}
\frac{[x_l - \mu_1 + 3\gamma]}{ [x_l - \mu_1 + \gamma]} \prod_{k=1}^L \frac{[x_l - \mu_k] [x_m - \mu_k + \gamma]}{[x_l - \mu_k + \gamma]} \prod_{\substack{k=1 \\ k \neq l, m}}^L \frac{[x_l - x_k + \gamma]}{[x_l - x_k]} \\
\hfill - \frac{[x_m - \mu_1 + 3\gamma]}{ [x_m - \mu_1 + \gamma]}  \prod_{k=1}^L [x_m - \mu_k] \prod_{\substack{k=1 \\ k \neq l, m}}^L \frac{[x_m - x_k + \gamma]}{[x_m - x_k]} \qquad l=r, m=s \nonumber \\
\frac{[\gamma] [x_m - x_s + \tau + (L+1)\gamma] [x_s - \mu_1 + 3\gamma]}{[\tau + (L+1)\gamma] [x_m - x_s] [x_s - \mu_1 + \gamma]} \prod_{k=1}^L [x_s - \mu_k] \prod_{\substack{k=1 \\ k \neq l, m, s}}^L \frac{[x_s - x_k + \gamma]}{[x_s - x_k]} \hfill l=r, m \neq s \nonumber \\
\frac{[\gamma] [x_m - x_r + \tau + (L+1)\gamma] [x_r - \mu_1 + 3\gamma]}{[\tau + (L+1)\gamma] [x_m - x_r] [x_r - \mu_1 + \gamma]} \prod_{k=1}^L [x_r - \mu_k] \prod_{\substack{k=1 \\ k \neq l, m, r}}^L \frac{[x_r - x_k + \gamma]}{[x_r - x_k]} \hfill l=s, m \neq r \nonumber \\
\frac{[\gamma] [x_l - x_s + \tau + (L+1)\gamma] [x_s - \mu_1 + 3\gamma]}{[\tau + (L+1)\gamma] [x_s - x_l] [x_s - \mu_1 + \gamma]} \prod_{k=1}^L \frac{[x_s - \mu_k] [x_m - \mu_k + \gamma]}{[x_l - \mu_k + \gamma]} \prod_{\substack{k=1 \\ k \neq l, m, s}}^L \frac{[x_s - x_k + \gamma]}{[x_s - x_k]}  \qquad m=r, l \neq s \nonumber \\
\frac{[\gamma] [x_l - x_r + \tau + (L+1)\gamma] [x_r - \mu_1 + 3\gamma]}{[\tau + (L+1)\gamma] [x_r - x_l] [x_r - \mu_1 + \gamma]} \prod_{k=1}^L \frac{[x_r - \mu_k] [x_m - \mu_k + \gamma]}{[x_l - \mu_k + \gamma]} \prod_{\substack{k=1 \\ k \neq l, m, r}}^L \frac{[x_r - x_k + \gamma]}{[x_r - x_k]}  \qquad m=s, l \neq r \nonumber \\
0 \hfill \text{otherwise}
\end{cases} . \\
\end{eqnarray}
\end{widetext}
The set of relations \eqref{Z}-\eqref{K} defines an explicit single determinant formula for the partition function \eqref{PF}.
Although the entries of $\Omega_{red}$ consist of straightforward simplifications of \eqref{FFbG}-\eqref{K}, it is worth remarking that
for practical computations it might be more convenient to write $\text{det} \left( \Omega \; \Omega_{red}^{-1} \right)$ as the ratio
$\text{det} \left( \Omega \right) / \text{det} \left( \Omega_{red} \right)$. In this way one avoids computing the inverse of $\Omega_{red}$
and evaluating the product $\Omega \; \Omega_{red}^{-1}$.

\section{Concluding remarks} \label{CONCL}

In the present paper we have obtained a novel representation for the partition function of the elliptic \textsc{sos} model in terms of a single
determinant. This result addresses a long standing question in the field and confirms the existence of such representations. In the limit 
$p \to 0$, where $[x]$ degenerates into a trigonometric function, followed by the limit $\tau \to \infty$; the partition function \eqref{PF}
reduces to that of the six-vertex model with domain-wall boundaries studied in \cite{Korepin_1982, Izergin_1987}. In contrast to the $L \times L$
matrix determinant representation found in \cite{Izergin_1987}, our solution consists of a determinant of a $L(L+3)/2 \times L(L+3)/2 $ matrix.
However, it is also important to notice that the determinant of \cite{Izergin_1987} is taken over a full-matrix, while in our case we have
a sparse matrix. In this way one might expect that \eqref{Z} is still liable to simplifications. Another interesting aspect of the representation
\eqref{Z} is related to the possibility of taking the homogeneous limit. In our case the partial homogeneous 
limit $\mu_i \to \mu$ can be obtained trivially in contrast to Izergin's representation for the six-vertex model.

Here we have singled out one particular possibility of determinant representation originated from the algebraic-functional framework. Alternative determinantal representations
are also possible and we plan to investigate them in a future publication \cite{Galleas_tba}. Moreover, it is quite remarkable the similarity between the 
roles played by the variables $x_0$ and $x_{\bar{0}}$ here and the mechanism employed in the localization method for the evaluation of path integrals 
\cite{Szabo_book}. This point certainly deserves further studies and we hope to address it in a future publication.

\appendix
\section{Numerical checks}

From definition \eqref{PF} we find $Z_{\tau} = [\gamma] [\tau + \gamma - \mu_1 + x_1]/[\tau + \gamma]$ for $L=1$. This is precisely the result obtained
from our representation \eqref{Z} with the help of summation formulae for Jacobi theta-functions. For $L > 1$ we can easily compare numerically the value of the 
partition function computed from the definition \eqref{PF} with the one obtained from our representation \eqref{Z}. This provides extra support for the 
validity of our results. Numerical evaluations have been performed with \texttt{Mathematica} and in Table \ref{tab: sets} one can find two sets
of randomly chosen values for the model's parameters. Tables \ref{tab: set1} and \ref{tab: set2} contain numerical comparisons using Set $1$  
and Set $2$ respectively for $2 \leq L \leq 5$.

\begin{table}
\caption{\label{tab: sets} Two sets of numerical values for the parameters.}
\begin{ruledtabular}
\begin{tabular}{lll}
Parameter & Set $1$ & Set $2$ \\ \hline
$x_1$;\;$\mu_1$ & $0.4327$;\;$0.6745$ & $0.8919$;\;$2.5449$ \\
$x_2$;\;$\mu_2$ & $1.0715$;\;$0.4129$ & $0.7233$;\;$1.8734$ \\
$x_3$;\;$\mu_3$ & $1.7481$;\;$3.3385$ & $0.1519$;\;$1.2745$ \\
$x_4$;\;$\mu_4$ & $2.2738$;\;$3.1245$ & $0.4388$;\;$2.0178$ \\
$x_5$;\;$\mu_5$ & $2.1415$;\;$1.9715$ & $2.6662$;\;$3.0089$ \\
$\gamma$ & $0.6512$ & $0.1219$ \\
$\tau$ & $0.1743$ & $0.2759$ \\
$p$ & $0.3116$ & $0.4421$ \\
\end{tabular}
\end{ruledtabular}
\end{table}

\begin{table}
\caption{\label{tab: set1} Numerical comparison using Set $1$.}
\begin{ruledtabular}
\begin{tabular}{lll}
$L$ & Definition \eqref{PF}  & Representation \eqref{Z} \\ \hline
$2$ & $0.00057111882715$ & $0.00057111882715$ \\
$3$ & $6.07562588434218 \; \ii$ & $6.07562588434047 \; \ii$ \\
$4$ & $6195.98835867588$ & $6195.98835851194$ \\
$5$ & $139.817171384552 \; \ii$ & $139.817171384640 \; \ii$ \\
\end{tabular}
\end{ruledtabular}
\end{table}

\begin{table}
\caption{\label{tab: set2} Numerical comparison using Set $2$.}
\begin{ruledtabular}
\begin{tabular}{lll}
$L$ & Definition \eqref{PF}  & Representation \eqref{Z} \\ \hline
$2$ & $0.230323036097808$ & $0.230323036097803$ \\
$3$ & $0.202679526300975 \; \ii$ & $0.202679526300981 \; \ii$ \\
$4$ & $2.659105034549285$ & $2.659105034415262$ \\
$5$ & $1478.397210835060 \; \ii$ & $1478.397210823134 \; \ii$ \\
\end{tabular}
\end{ruledtabular}
\end{table}

\begin{acknowledgments}
The author thanks H. Rosengren for valuable discussions and F. R\"uhle for help with numerical checks.
\end{acknowledgments}

\bibliography{references1.bib}

\end{document}